\documentclass[prd,twocolumn,nofootinbib,showpacs,superscriptaddress]{revtex4}
\usepackage{amsmath}
\usepackage{epsfig}
\begin{document}
\title{How to move a black hole without excision: gauge conditions for the numerical evolution of a moving puncture}

\author{James R. van Meter}
\affiliation{Gravitational Astrophysics Laboratory,
NASA Goddard Space Flight Center, 8800 Greenbelt Rd.,
Greenbelt, MD 20771, USA}
\author{John G. Baker}
\affiliation{Gravitational Astrophysics Laboratory,
NASA Goddard Space Flight Center, 8800 Greenbelt Rd.,
Greenbelt, MD 20771, USA}
\author{Michael Koppitz}
\affiliation{Gravitational Astrophysics Laboratory,
NASA Goddard Space Flight Center, 8800 Greenbelt Rd.,
Greenbelt, MD 20771, USA}
\author{Dae-Il Choi}
\affiliation{Gravitational Astrophysics Laboratory,
NASA Goddard Space Flight Center, 8800 Greenbelt Rd.,
Greenbelt, MD 20771, USA}
\affiliation{Universities Space Research Association,
10211 Wincopin Circle, Suite 500, Columbia, MD 21044}

\begin{abstract}
Recent demonstrations of unexcised black holes traversing 
across computational grids represent a significant advance in numerical
relativity.  Stable and accurate simulations of multiple orbits, and their
radiated waves, result.  This capability is critically undergirded by
a careful choice of gauge.  Here
we present analytic considerations which suggest certain gauge choices,
and numerically demonstrate their efficacy in evolving a single moving
puncture black hole. 
\end{abstract}
\maketitle
\section{Introduction}

Gravitational waveforms from binary black hole mergers can only be obtained
through 3D numerical relativity simulations of the full Einstein equations.
Such simulations have proven to be very challenging. Improvements have
often come from new formulations or a new choice of gauge. For example, a
corotating gauge led to the first evolution of a binary through one
full orbit~\cite{Bruegmann:2003aw} as well as through a plunge, merger, and
ringdown~\cite{Alcubierre:2004hr}.

Recently, the authors of this paper~\cite{Baker:2005vv} and another research
group~\cite{Campanelli:2005dd} independently developed the capability to
numerically evolve freely moving black holes, i.e. black holes that are in
no way fixed to any position in the computational grid, without using excision techniques for the black hole interiors.  These efforts led to
the simulation of record-breaking numbers of orbits, among other 
accomplishments \cite{Baker:2006yw}. Both of these research programs employed gauge 
conditions that differed not only from those
prescribed in prior literature, but also from each other.  A careful choice
of gauge is important, particularly for moving black holes, as it can
determine whether a simulation is stable or becomes irrevocably corrupted by
a runaway build up of numerical error.  There are various ways in which a
choice of gauge can affect the stability and accuracy of a numerical
simulation; for example, a poor choice of gauge might cause large gradients
in the fields which in turn engender large finite differencing error. 

In this paper we investigate a group of gauge choices analytically, and
numerically demonstrate their effects on moving black hole simulations. These 
gauge choices are framed in the context of a conventional, 3+1,
conformal formulation of Einstein's equations known as the 
Baumgarte-Shapiro-Shibata-Nakamura (BSSN) formulation
\cite{Nakamura87,Shibata95,Baumgarte99,Imbiriba:2004tp}.
After a brief exploration of the shortcomings of a commonly used
gauge choice in ~\ref{sec:orig_gauge}, we calculate in
section~\ref{sec:eigenvalues} the characteristic speeds for various gauges
and then show the numerical behavior of some of them in section~\ref{sec:numrel}.
In section~\ref{sec:properties} we discuss the properties of the most promising gauge choices and 
suggest further improvements.  Conclusions are presented in section~\ref{sec:conclusions}.

\section{Original 1+log slicing and Gamma-driver shift conditions}
\label{sec:orig_gauge}
We consider first a gauge introduced in \cite{Alcubierre02a} for evolving non-moving punctures
without excision: 

\begin{eqnarray}
\partial_t\alpha &=& -2\alpha K \label{eqn:original_a}\\
\partial_t\beta^i&=& \frac{3}{4}\alpha \Psi_0^{-n}B^i \label{eqn:original_b}\\
\partial_tB^i    &=& \partial_t{\tilde \Gamma}^i-\eta B^i \label{eqn:original_c}
\end{eqnarray}
where $\alpha$ is the lapse, $\beta^i$ is the shift, $K$ and ${\tilde \Gamma}^i$ are the usual BSSN variables, 
$\Psi_0^4$ is the initial conformal factor ($\Psi_{BL}^4$ in
\cite{Alcubierre02a}), 
$B^i$ is an auxiliary variable to make the shift equation first order in time, and $\eta$ is a constant damping factor.
This gauge has proven to be very effective at evolving a single non-moving puncture 
for arbitrarily long durations, and in some circumstances it can also be applied to
binary punctures.  Here the condition on the lapse, a variation of the Bona-Masso slicing condition \cite{Bona:1994dr,Alcubierre02a}, 
is the standard ``1+log" singularity-avoiding slicing condition.  Of particular interest is the ``Gamma-driver" condition 
for the shift, which is designed to ultimately ``freeze Gamma", i.e. drive $\partial_t{\tilde \Gamma}$ to zero as 
the physics of the spacetime also evolve towards quiescence.

For $n>0$, $\Psi_0^{-n}=0$ at the puncture, and thus this specific gauge will ensure that $\beta^i$ will not evolve there.
As the motion of the puncture is determined by the shift, the puncture will remain motionless in this gauge 
since the shift is initially zero.  For black holes that are physically dynamical the coordinates may become highly distorted.  
For example, in the case of binary black holes, the separation 
between the black holes as determined by the proper distance from horizon to horizon in some spatial slice
must decrease with time.  This implies that components of the 3-metric will approach zero between the black holes.  
Then the evolved part of the BSSN conformal exponent $\phi$ must either approach $-\infty$, or components of the conformal
3-metric ${\tilde \gamma}_{ij}$ must approach zero.  In the latter case the inverse 3-metric must diverge in order to maintain a unit determinant.
Thus, in the fixed puncture approach, arbitrarily large, ever-increasing fields and gradients around the black holes seem likely
to develop and cause numerical difficulties.  Indeed this can be demonstrated even in the case of a single moving black hole.  

We will use as a test case a single black hole, given by Bowen-York puncture initial data \cite{Brandt97b} (as computed by the elliptic solver {\tt AMRMG} \cite{Brown:2004ma}), with unit puncture mass and momentum such that 
its physical speed should be half the speed of light.  We evolve this data with our {\tt Hahndol}
code\cite{Imbiriba:2004tp}, which uses the usual, conformal BSSN formulation of Einstein's evolution equations on a cell-centered
numerical grid, with $4^{\rm th}$-order spatial differencing and $4^{\rm th}$-order Runge-Kutta time-integration.
The initial puncture position is at coordinates $(-3M,0,0)$ (where $M$ is the puncture mass) and the momentum is in
the positive $x$-direction.  The $x$-axis is between grid points, which are a distanced $M/16$ apart; all data presented here has been interpolated onto the $x$-axis for plotting
purposes.
$\beta^i$ and $B^i$ are initialized to zero, while the lapse is precollapsed with the profile $\Psi_0^{-2}$ as 
suggested in \cite{Alcubierre02a} (and recommended in \cite{Campanelli:2005dd} for moving punctures).  The damping parameter $\eta$ is $2$, unless otherwise stated.  
We will assess the performance of each gauge by the evolving behavior of the 
quantities $\alpha$, $\beta^x$, $\phi$, and ${\tilde \Gamma^x}$;
in particular the peak in $\phi$ will roughly indicate the position of the puncture and any extreme gradients in ${\tilde \Gamma^x}=-\partial_j{\tilde \gamma}^{jx}$ will
tend to indicate problems with the gauge.

In this regard it should be noted that ${\tilde \Gamma^i}$ is completely controllable by the gauge condition, and so any undesirable features observed in ${\tilde \Gamma^i}$
are in principle avoidable via a better choice of gauge.  In particular, the specific ``Gamma-freezing" condition ${\tilde \Gamma^i}=0$ may be desirable, as this elliptic
generalization of isotropic coordinates proposed by Dirac is expected to be non-pathological\cite{Dirac:1959,Smarr:1978}.  So, among
hyperbolic Gamma-driving conditions, those that result in smaller values of ${\tilde \Gamma^i}$ and its derivatives might be preferred.

\begin{figure}
\includegraphics[scale=.36, angle=-90]{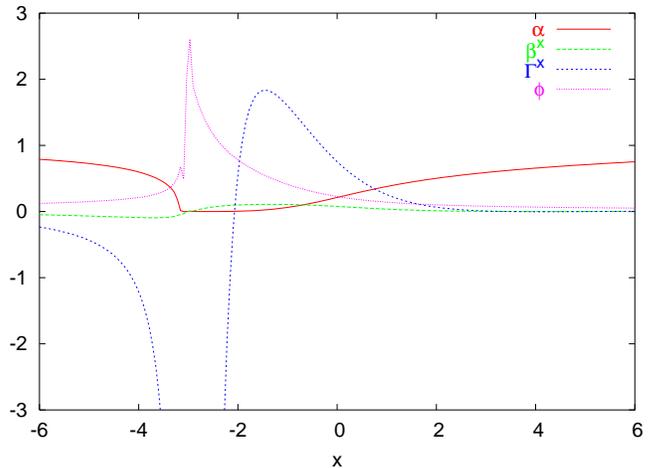}
\caption{Results of the non-moving puncture gauge $\partial_t\alpha  =  -2\alpha K$,
$\partial_t\beta^i =  \frac{3}{4}\alpha \Psi_0^{-n}B^i$, 
$\partial_tB^i     =  {\dot{\tilde \Gamma}^i}-\eta B^i$ at time $t=17M$.
The conformal metric becomes singular by $t=18M$.}
\label{gauge_original}
\end{figure}

In Fig.~\ref{gauge_original}, evolving with the original gauge given in Eqs.~(\ref{eqn:original_a}-\ref{eqn:original_c}), we see that ${\tilde \Gamma^x}$ grows very large and $\phi$ develops sharp features.  Shortly thereafter 
the inverse conformal metric diverges, which is our criterion for stopping the run.  This failure motivates investigation of moving punctures.

\begin{figure}
\includegraphics[scale=.36, angle=-90]{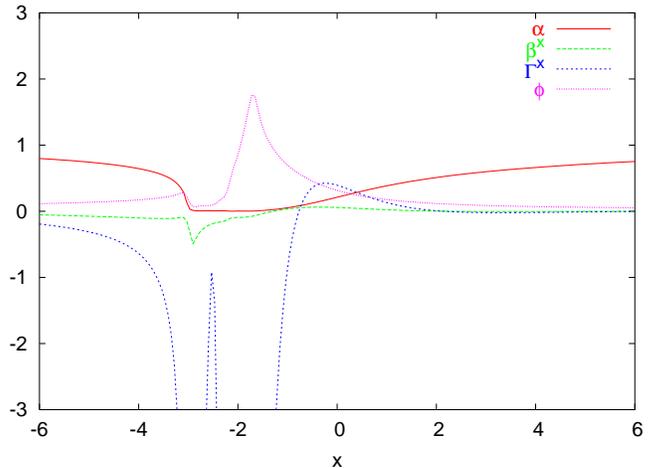}
\caption{Results of the gauge $\partial_t\alpha  =  -2\alpha K$,
$\partial_t\beta^i =  \frac{3}{4}\alpha B^i$,
$\partial_tB^i     =  {\dot{\tilde \Gamma}^i}-\eta B^i$ at time $t=17M$.
Sharp features around $x=-3M$ fail to propagate.
The conformal metric becomes singular by $t=18M$.}
\label{gp_0000_lapsed}
\end{figure}

If we wish the punctures to move we must allow $\beta^i$ to
evolve freely away from zero and thus we must take the 
conformal factor out of the shift equation (setting $n=0$ in Eq.~(\ref{eqn:original_b})).  The results, shown in Fig.~\ref{gp_0000_lapsed},
are not much improved.  The puncture is now free to move but non-propagating features
at $x=-3M$ destabilize the simulation.  Clearly there is a problem with zero-speed modes.  The extended, collapsed region of the lapse 
points to a particular difficulty in this regard, and suggests an improvement.

\begin{figure}
\includegraphics[scale=.36, angle=-90]{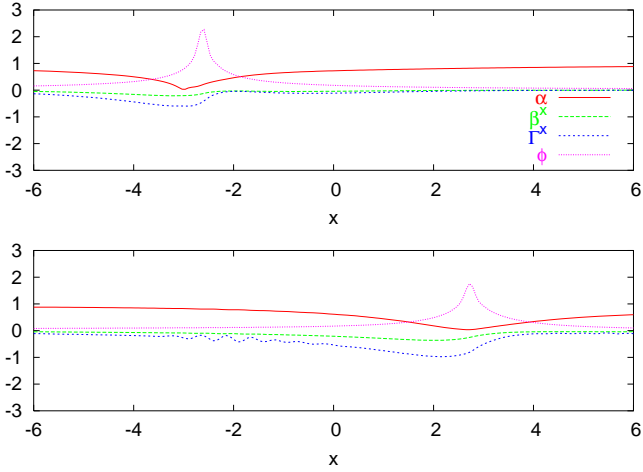}
\caption{Results of the gauge $\partial_t\alpha  =  -2\alpha K$,
$\partial_t\beta^i =  \frac{3}{4}B^i$, 
$\partial_tB^i     =  {\dot{\tilde \Gamma}^i}-\eta B^i$ at times $t=4.5M$ (top panel) and $t=30M$ (bottom panel).  The top panel clearly shows the lapse
lagging behind the puncture, while the bottom panel exhibits noise
propagating from the puncture.
}
\label{gauge_0000}
\end{figure}

The factor of $\alpha$ in Eq.~(\ref{eqn:original_b}) should also
be removed, for even if it were not precollapsed it would still be driven nearly to zero
around the puncture by the 1+log slicing condition, and therefore may retard the evolution of 
$\beta^i$.  When this factor is removed, the results are not as catastrophic as before, but the simulation is plagued with noise.  Fig.~\ref{gauge_0000} shows, in
particular, a tendency for the minimum in the lapse to lag behind the puncture position.  The above results suggest that a careful study of the 
propagation speeds in this system might be helpful.

\section{Characteristic speeds}
\subsection{Eigenvalue analysis}
\label{sec:eigenvalues}
A potential danger of zero-speed modes is that errors may couple to them, compound in place, and become inordinantly large.
Zero-speed modes seem particularly hazardous in the case of moving black holes as they may adversely affect the dynamics by putting a drag
on the puncture motion.  In some cases zero-speed modes can also be a source of instability \cite{Alcubierre99e}.  So there is good reason
to be aware of such modes.

For the purpose of investigating characteristic speeds,
a simple yet revealing method of analysis is to linearize the equations, assume plane wave solutions, and solve for the
eigenmodes of the resulting algebraic system.
With the gauge conditions described above, the BSSN equations can be linearized about flat space as follows.

\begin{eqnarray}
{\dot a} &=& -2\alpha_0K\\
{\dot B^i} &=& {\dot{\tilde \Gamma}^i}\\
{\dot b^i} &=& \frac{3}{4}\alpha_0^p B\\
{\dot \phi} &=& -{1\over 6}(\alpha_0K - \partial_i b^i)+\beta_0^k\partial_k\phi\\
{\dot K} &=& - \partial_i \partial_i a +\beta_0^k\partial_k K\\
{\dot h}_{ij} &=& -2 \alpha_0{\tilde A}_{ij} + \partial_i b^j + \partial_jb^i - {2\over 3}\delta_{ij} \partial_kb^k +\beta_0^k\partial_k h_{ij}\\
{\dot{\tilde A}}_{ij} &=& \left[ -\partial_i\partial_ja - {1\over 2}\alpha_0\partial_k\partial_kh_{ij} + {1\over 2}\alpha_0 \partial_i{\tilde\Gamma}^j + {1\over 2}\alpha_0 \partial_j{\tilde\Gamma}^i\right. \nonumber \\ && \left.- 2\alpha_0\partial_i \partial_j\phi \right]^{\rm TF} +\beta_0^k\partial_k {\tilde A}_{ij}\\
\dot{\tilde\Gamma}^i &=& -{4\over 3} \alpha_0\partial_i K + \partial_k\partial_k b^i + {1\over 3} \partial_i \partial_j b^j +\beta_0^k\partial_k {\tilde\Gamma}^i
\end{eqnarray}
where $a \equiv \alpha-\alpha_0$, $b^i \equiv \beta^i-\beta_0^i$, $h_{ij} \equiv {\tilde \gamma_{ij}}-\delta_{ij}$, and the damping term on $B^i$ has been dropped as it
should not affect the real part of the characteristic speeds.  The lapse and shift are assumed to have
constant, zeroth order terms $\alpha_0$ and $\beta_0^i$, respectively which conveniently allows us to capture 
essential effects of
the lapse factors and the advection terms without making the eigenvalue problem intractable.

For spatial variation only in one dimension, with no initial transverse components, and assuming plane-wave solutions, the
above system of equations can be written in the form:
\begin{eqnarray}
\label{ket_evolution}
\partial_t|u\rangle = ik\mathsf{M}|u\rangle 
\end{eqnarray}
where
\begin{equation}
|u\rangle = \left( \begin{array}{c} {\hat a} \\ {\hat B} \\ {\hat b} \\ {\hat \phi} \\ {\hat K} \\ {\hat h} \\ {\hat A} \\ {\hat \Gamma}   \end{array}
\right)e^{i(kx-\omega t)} 
\end{equation}
with ${\hat a}$, ${\hat B}$, ${\hat b}$, ${\hat \phi}$, ${\hat K}$, ${\hat h}$, ${\hat A}$, and ${\hat \Gamma}$ the constant amplitudes of $a$, $B^x$, $b^x$, $\phi$,
$K$, $h_{xx}$, ${\tilde A}_{xx}$, and ${\tilde \Gamma}^x$ respectively,
and $\mathsf{M}=$
\begin{equation}
 \left( \begin{array}{c c c c c c c c} 
       0         &        0        &         0       &       0       & -\frac{2}{ik}\alpha_0 &       0       &       0       &       0       \\
       0         &        0        &   \frac{4ik}{3} &       0       &  \frac{4}{3}\alpha_0  &       0       &       0       &-\beta_0       \\
       0         & \frac{3\alpha_0^p}{4ik}   &         0       &       0       &       0       &       0       &       0       &       0       \\
       0         &        0        &  -\frac{1}{6}   &  -\beta_0     & -\frac{1}{6ik}\alpha_0 &       0       &       0       &       0       \\
       -ik       &        0        &         0       &       0       &  -\beta_0     &       0       &       0       &       0       \\
       0         &        0        &   -\frac{4}{3}  &       0       &        0      &   -\beta_0    & -\frac{2}{ik}\alpha_0 &       0       \\
 -\frac{2ik}{3}  &        0        &         0       & -\frac{4ik}{3}\alpha_0 &        0      & -\frac{ik}{2}\alpha_0 &  -\beta_0     & -\frac{2}{3}\alpha_0  \\
       0         &        0        & \frac{4ik}{3}   &       0       &  \frac{4}{3}\alpha_0  &       0       &       0       &   -\beta_0    
\end{array} \right)            
\end{equation}


The eigenvalues of $\mathsf{M}$ are
$0$,$-\beta_0$,$-\beta_0+\alpha_0$,$-\beta_0-\alpha_0$,$-\frac{1}{2}\beta_0+\frac{1}{2}\sqrt{\beta_0^2+8\alpha_0}$,$-\frac{1}{2}\beta_0-\frac{1}{2}\sqrt{\beta_0^2+8\alpha_0}$,$-\frac{1}{2}\beta_0+\frac{1}{2}\sqrt{\beta_0^2+4\alpha_0^p}$,
and $-\frac{1}{2}\beta_0-\frac{1}{2}\sqrt{\beta_0^2+4\alpha_0^p}$. 
If $|v \rangle$ is the eigenvector associated with eigenvalue $v$, and the covector $\langle v |$ is defined such that $\sum_{v} |v\rangle\langle
v| = \mathsf{I}$, 
then the eigen decomposition of $\mathsf{M}$ allows 
the system of evolution equations to be written as a series of advection terms, each
associated with a characteristic
velocity equal to one of the eigenvalues:  
\begin{eqnarray}
\label{eqn:linBSSNadvectionResult}
 \partial_t |u\rangle  &=& \sum_{v} v \partial_x|v\rangle \langle v |u\rangle
\end{eqnarray}
The inner product with $\langle v|$ gives:
\begin{eqnarray}
\label{eqn:eigenfield}
 \partial_t \langle v |u\rangle  &=& v \partial_x\langle v |u\rangle
\end{eqnarray}

Now we are interested in slow speed modes.  Note that although $\beta^i$ is typically initialized at zero, with the Gamma-driver condition it evolves rapidly
and significantly enough from zero that we do not classify $\beta_0$ as a slow mode (since we're interested in moving punctures, $\beta^i$ is generally non-vanishing).
In addition to the obvious zero speed mode there are two modes that approach zero in the limit where $\alpha_0$ approaches zero.  
If $|\alpha_0|<<|\beta_0|$      
then $-\frac{1}{2}\beta_0+\frac{1}{2}\sqrt{\beta_0^2+8\alpha_0}$ and $-\frac{1}{2}\beta_0+\frac{1}{2}\sqrt{\beta_0^2+4\alpha_0^p}$
become approximately $4\alpha_0/\beta_0$ and $2\alpha_0^p/\beta_0$, respectively.  In the full nonlinear system, the 1+log 
slicing condition will collapse $\alpha$ nearly to zero in a finite region around the puncture (where $\beta^i$ will be nonzero for a moving puncture).
Modes which propagate at a speed approximately proportional to a positive power of $\alpha$ for small $\alpha$, which we will refer to as $\alpha$-speed modes,
are thus of potential concern.  In particular, late in the evolution of black hole binaries there will emerge a significant, stationary region in which
the lapse has collapsed nearly to zero, in which $\alpha$-speed modes are effectively zero-speed modes.

The eigencovector for $v=-\frac{1}{2}\beta_0+\frac{1}{2}\sqrt{\beta_0^2+8\alpha_0}$ involves only the fields $a$ and $K$, is relatively independent from the
other fields, and easy to fix, so we will address it first.  This $\alpha$-speed
mode can be understood also in the context of the original nonlinear equations by noting that $\alpha$ couples with $K$ to give a wave equation
such that the speed of fluctuations of the lapse goes to zero when the lapse itself goes to zero.  It is particularly egregious when the lapse couples
to an $\alpha$-speed mode, as once it collapses nearly to zero it will tend to get ``stuck" there.  But this can be remedied in a natural way simply by adding
an advection term as follows,
\begin{equation}
  {\dot \alpha} = -2\alpha K + \beta^j\partial_j\alpha,
\label{eqn:shiftinglapse}
\end{equation}
which will modify the principal part of its wave equation such that when $\alpha$ goes to zero its
speed will go to $\beta$.  Note that Eq.~(\ref{eqn:shiftinglapse}), first used for moving punctures in \cite{Campanelli:2005dd}, is consistent with the original Bona-Mass family of slicing conditions\cite{Bona:1994dr}.

The eigencovector for $-\frac{1}{2}\beta_0+\frac{1}{2}\sqrt{\beta_0^2+4\alpha_0^p}$ is more complicated, involving not only $a$ but several other fields as well.  It turns
out that advecting the lapse will not remove this mode.  But noting 
that this eigenfield involves $b^x$, and that the equation for $b^x$ is conspicuously absent an advection term, an obvious stratagem is to advect the original $\beta^i$ as follows:
\begin{eqnarray}
{\dot \beta^i} &=& \frac{3}{4}B^i + \beta^j\partial_j\beta^i
\label{eqn:shiftingshift}
\end{eqnarray}

Finally, Eq.~(\ref{eqn:eigenfield}) for the eigenvalue $v=0$ yields 
\begin{eqnarray}
\partial_t (B^x-{\tilde \Gamma}^x) = 0
\end{eqnarray}
This equation, a trivial consequence of the original Gamma-driver condition,
immediately suggests a simple modification.  The most natural way to remove this zero-speed mode is to advect 
it at $\beta$ speed by modifying the ${\dot B^i}$ equation thus:
\begin{eqnarray}
\partial_t B^i = \partial_t {\tilde \Gamma}^i+\beta^j\partial_j (B^i-{\tilde\Gamma}^i) - \eta B^i
\label{eqn:shiftingBminusGamma}
\end{eqnarray}
It turns out that the $+\beta^j\partial_j B^i$ term alone will not suffice, as for example it might introduce an exponentially growing mode.  Meanwhile the
$-\beta^j\partial_j {\tilde\Gamma}^i$ term alone adds another $\alpha$-speed mode to the system.

The foregoing analysis suggests the addition of four advection terms to the gauge equations, as indicated in
Eqs.~(\ref{eqn:shiftinglapse},\ref{eqn:shiftingshift},\ref{eqn:shiftingBminusGamma}).  The lapse equation is relatively independent of the others: if it is advected then a 
particular $\alpha$-speed mode is removed, and if it is not advected then this particular $\alpha$-speed mode remains.
But the three advection terms suggested for the shift equations are strongly interdependent, so their combined effects are not so obvious and it is instructive to 
consider which undesirable modes result from which
combinations of these three terms.  The possible combinations and their resulting ``bad" speeds are summarized in Table 1.  Here we see that for $p>0$, only when all the above 
advection terms are included will the system be free of slow-speed
and exponentially growing modes; while for $p=0$, the $-\beta^j\partial_j{\tilde \Gamma}^i$ term remains critical.  In the case with all advection terms the benign eigenspeeds are:  $-\beta_0$, $-\beta_0$, $-\beta_0-\alpha_0$,
$-\beta_0+\alpha_0$, $-\beta_0-\sqrt{\alpha_0^p}$,  $-\beta_0+\sqrt{\alpha_0^p}$, $-\beta_0-\sqrt{2\alpha_0}$,  $-\beta_0+\sqrt{2\alpha_0}$ 
(where $-\beta_0$ is listed twice because it is now associated with two distinct eigenvectors).  

\bigskip
\begin{table}[h]
\begin{center}
\begin{tabular}{|c|c|c|c|l|}
\hline
Case \# & $+\beta^j\partial_j B^i$ & $-\beta^j\partial_j
{\tilde\Gamma}^i$ & $+\beta^j\partial_j \beta^i$ & ``bad" speeds
\\
\hline
\hline
1 & N & N & N & 0, $\alpha^p$ \\
\hline
2 & N & N & Y & 0 \\
\hline
3 & N & Y & N & $\alpha^{p/2}$ \\
\hline
4 & N & Y & Y & $\alpha^p$ \\
\hline
5 & Y & N & N & 0 \\
\hline
6 & Y & N & Y & $-i$ \\
\hline
7 & Y & Y & N & $\alpha^p$  \\
\hline
8 & Y & Y & Y & none \\
\hline
\end{tabular}
\end{center}
\caption{Effect of various advection terms in the shift equations on the presence of undesirable eigenspeeds.  'Y' or 'N' indicates whether each advection term is
included or
not, respectively, in Eq.~(\ref{eqn:shiftingshift}) or Eq.~(\ref{eqn:shiftingBminusGamma}).
0, $\alpha^p$, or $-i$ indicates whether the resulting linearized equations have a zero-speed mode, an $\alpha$-speed mode, or
an exponentially growing mode, respectively.  The lapse is assumed to be advected as in Eq.~(\ref{eqn:shiftinglapse}); otherwise an additional $\alpha$-speed
mode would appear in every case.
}
\label{table}
\end{table}
\bigskip

\subsection{Numerical tests}
\label{sec:numrel}

Evidence of the slow-speed modes found in the linearized analysis can often (but not always) be found in simulations of the full nonlinear system.
Note that zero-speed modes only represent a potential danger.  In some cases errors may not couple to these modes even though they exist in the equations.

\begin{figure}
\includegraphics[scale=.36, angle=-90]{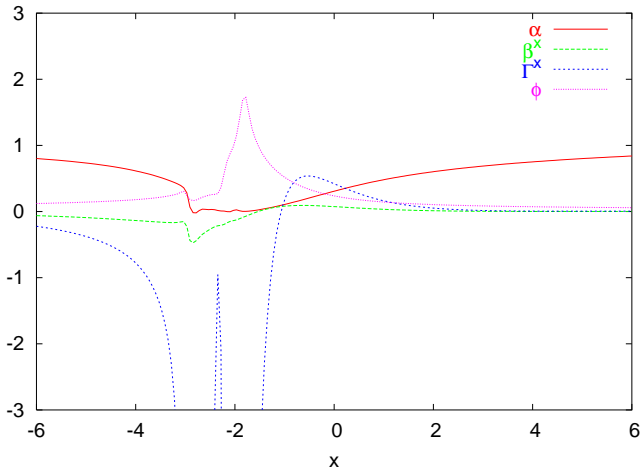}
\caption{Results for the gauge ${\partial_t \alpha} = -2\alpha K + \beta^j\partial_j\alpha $, 
${\partial_t \beta^i} = \frac{3}{4}\alpha B^i$,
${\partial_t B^i} = {\partial_t{\tilde \Gamma}^i} 
-\eta B^i $ (Case \#1 in Table~\ref{table} with p=1) at time $t=14M$.  Non-propagating features are evident around $x=-3M$.  The conformal metric becomes singular by $t=15M$
}
\label{gp_1000_lapsed}
\end{figure}

\begin{figure}
\includegraphics[scale=.36, angle=-90]{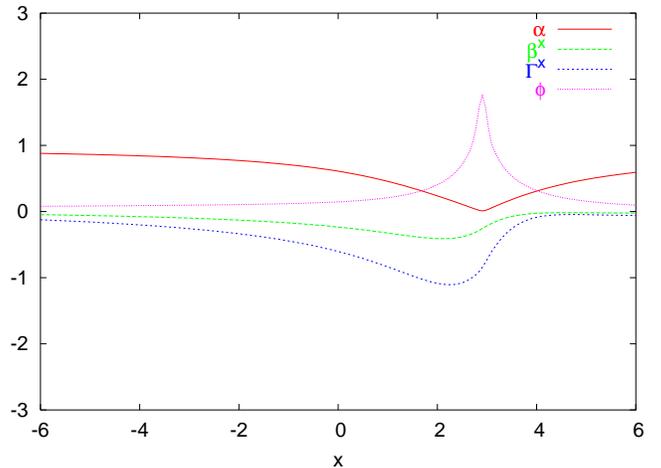}
\caption{Results for the gauge ${\partial_t \alpha} = -2\alpha K + \beta^j\partial_j\alpha $,
${\partial_t \beta^i} = \frac{3}{4}B^i$,
${\partial_t B^i} = {\partial_t{\tilde \Gamma}^i} 
-\eta B^i $ (Case \#1 in Table~\ref{table} with p=0) at time $t=30M$.  
The evolution appears smooth.
}
\label{gp_1000}
\end{figure}

\begin{figure}
\includegraphics[scale=.36, angle=-90]{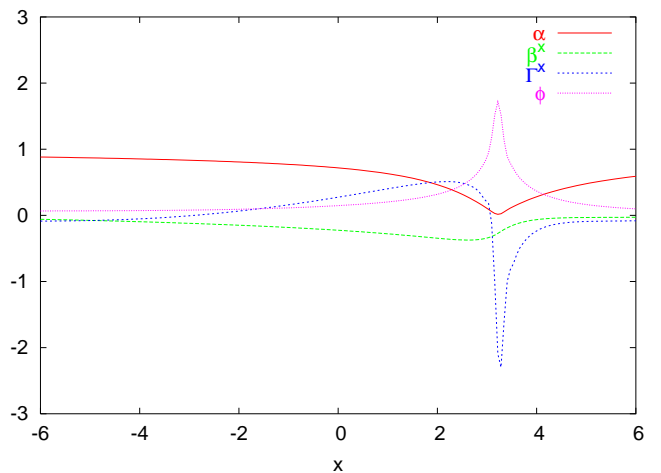}
\caption{Results for the gauge ${\partial_t \alpha} = -2\alpha K +
\beta^j\partial_j\alpha $,
${\partial_t \beta^i} = \frac{3}{4}\alpha B^i $,
${\partial_t B^i} = {\partial_t{\tilde \Gamma}^i} -
\beta^j\partial_j{\tilde \Gamma}^i
-\eta B^i $ (Case \#3 in Table~\ref{table} with p=0) at time $t=30M$.
${\tilde \Gamma}^x$ continues to grow larger and its gradient steeper.
}
\label{gp_1010}
\end{figure}

\begin{figure}
\includegraphics[scale=.36, angle=-90]{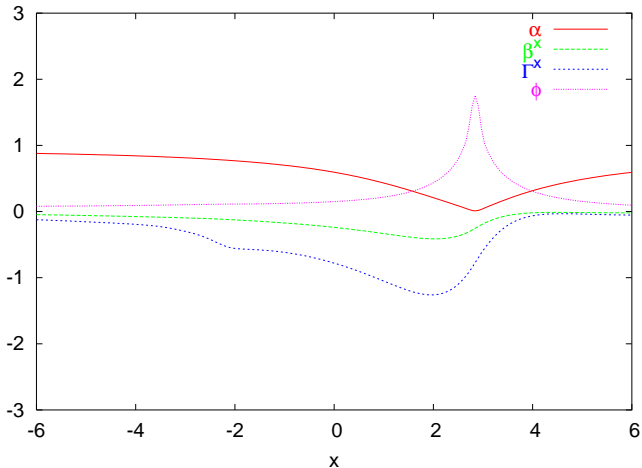}
\caption{Results for the gauge ${\partial_t \alpha} = -2\alpha K + \beta^j\partial_j\alpha $,
${\partial_t \beta^i} = \frac{3}{4}\alpha B^i $,
${\partial_t B^i} = {\partial_t{\tilde \Gamma}^i} + \beta^j\partial_jB^i
-\eta B^i $ (Case \#5 in Table~\ref{table} with p=0) at time $t=30M$.
The feature around $x=-2M$ does not propagate.
}
\label{gp_1100}
\end{figure}

\begin{figure}
\includegraphics[scale=.36, angle=-90]{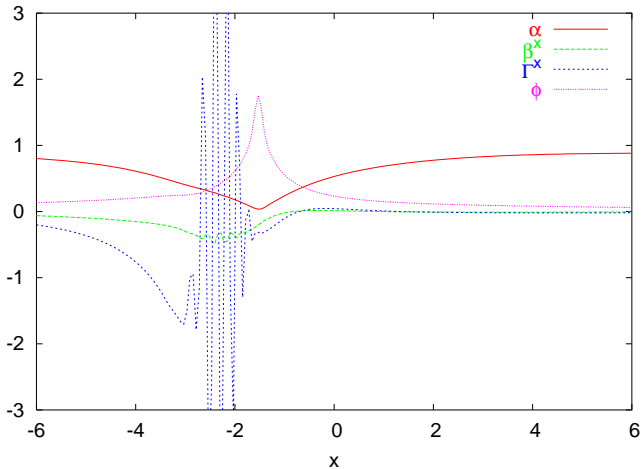}
\caption{Results for the gauge ${\partial_t \alpha} = -2\alpha K + \beta^j\partial_j\alpha $,
${\partial_t \beta^i} = \frac{3}{4}\alpha B^i + \beta^j\partial_j\beta^i$,
${\partial_t B^i} = {\partial_t{\tilde \Gamma}^i} + \beta^j\partial_jB^i -\eta B^i $ 
(Case \#6 in Table~\ref{table} with p=0) at time $t=10M$  
The noise in ${\tilde \Gamma}^x$ grows exponentially.
The conformal metric becomes singular by $t=11M$.
}
\label{gp_1101}
\end{figure}

\begin{figure}
\includegraphics[scale=.36, angle=-90]{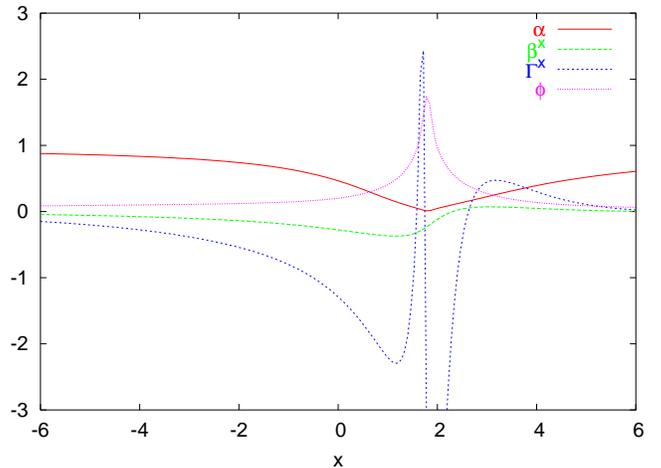}
\caption{Results for the gauge ${\partial_t \alpha} = -2\alpha K + \beta^j\partial_j\alpha $,
${\partial_t \beta^i} = \frac{3}{4}\alpha B^i + \beta^j\partial_j\beta^i$,
${\partial_t B^i} = {\partial_t{\tilde \Gamma}^i} + \beta^j\partial_j(B^i-{\tilde \Gamma}^i) -\eta B^i $ 
(Case \#8 in Table~\ref{table} with p=1) at time $t=30M$.
Aside from the sharp features in ${\tilde \Gamma}^i$, which do not grow 
significantly in time, the propagation is non-pathological.
}
\label{gp_1111_lapsed}
\end{figure}

\begin{figure}
\includegraphics[scale=.36, angle=-90]{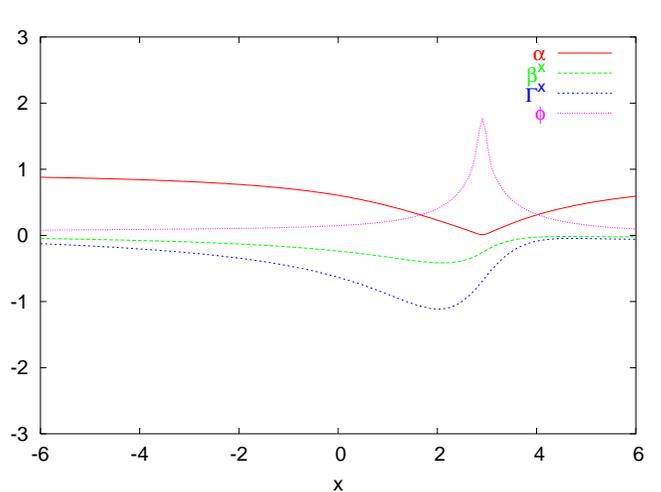}
\caption{Results for the gauge ${\partial_t \alpha} = -2\alpha K + \beta^j\partial_j\alpha $,
${\partial_t \beta^i} = \frac{3}{4}B^i + \beta^j\partial_j\beta^i$,
${\partial_t B^i} = {\partial_t{\tilde \Gamma}^i} + \beta^j\partial_j(B^i-{\tilde \Gamma}^i) -\eta B^i $
(Case \#8 in Table~\ref{table} with p=0) at time $t=30M$.
The evolution is very smooth.
}
\label{gp_1111}
\end{figure}

Here we provide a few examples of the cases given in Table~\ref{table}.  Fig.~\ref{gp_1000_lapsed} represents Case \#1 with $p=1$, which is predicted
to have a zero-speed mode and indeed manifests a non-propagating feature at $x=-3M$.  On the other hand, Case \#1 with $p=0$, also predicted to have a 
zero-speed mode, belies no indication of it in Fig.~\ref{gp_1000},
instead appearing quite smooth.  In this case, evidently, the zero-speed mode does not couple significantly with the other fields.  
In Fig.~\ref{gp_1010}, 
the spike in ${\tilde \Gamma}^x$, which becomes steeper in time as the ``tail" in ${\tilde \Gamma}^x$ grows,
is apparently unrelated to slow-speed modes as none are predicted in this Case \#3 with $p=0$.
However, for a longer evolution this growing gradient in ${\tilde \Gamma}^x$ can be expected to
adversely affect the constraints.  In Fig.~\ref{gp_1100} the tail in ${\tilde \Gamma}^x$ and in particular the non-propagating
bend in the tail around $x=-2M$ seems to be evidence of the zero-speed mode predicted for Case \#5.  In Fig.~\ref{gp_1101}, noise in ${\tilde \Gamma}^x$ grows 
exponentially, as expected for Case \#6.  And finally, Case \#8, depicted in Fig.~\ref{gp_1111_lapsed} and Fig.~\ref{gp_1111}, is demonstrated to be free
of the previously identified ``bad" speed modes whether $p=0$ or $p=1$.  (Although the former appears much smoother, perhaps because the 
eigenspeeds $-\beta_0\pm 1$ allow faster propagation of error away from the puncture than $-\beta_0\pm\sqrt{\alpha_0}$.)

\section{Properties of the ``cleanest" gauges}
\label{sec:properties}
Of all the combinations suggested in Table~\ref{table}, two distinguish themselves in numerical tests as conducive
to particularly smooth propagation of the moving black hole.  These gauges are Case \#8 with $p=0$, which we will
refer to as the ``shifting-shift case":

\begin{eqnarray} 
{\partial_t \beta^i} - \beta^j\partial_j\beta^i &=& \frac{3}{4}B^i  \label{eqn:shiftingshift_a}\\
{\partial_t B^i} - \beta^j\partial_jB^i &=& {\partial_t{\tilde \Gamma}^i} -\beta^j\partial_j{\tilde \Gamma}^i -\eta B^i \label{eqn:shiftingshift_b} 
\end{eqnarray}

and Case \#1 with $p=0$, which we will refer to as the ``non-shifting-shift" case:

\begin{eqnarray} 
{\partial_t \beta^i} &=& \frac{3}{4}B^i \label{eqn:nonshiftingshift_a} \\ 
{\partial_t B^i} &=& {\partial_t{\tilde \Gamma}^i} -\eta B^i \label{eqn:nonshiftingshift_b} 
\end{eqnarray}

The efficacy of Eqs.~(\ref{eqn:nonshiftingshift_a}-\ref{eqn:nonshiftingshift_b}) in evolving moving punctures was originally demonstrated in \cite{Campanelli:2005dd,Campanelli:2006gf}.  Apparently 
the zero-speed mode is not problematical in this case.  Regarding the ``shifting-shift" condition,
Eqs.~(\ref{eqn:shiftingshift_a}-\ref{eqn:shiftingshift_b}), we have recently used it successfully in the stable evolution of two equal mass black holes through $~5.5$ orbits.
This latter gauge has also been recommended for the strong hyperbolicity it brings to the BSSN equations \cite{Gundlach:2006tw,Beyer:2004sv}, suggesting
it may be a more robustly stable choice than Eqs.~(\ref{eqn:nonshiftingshift_a}-\ref{eqn:nonshiftingshift_b}).  Nevertheless for the single black hole simulations 
represented by Figs.~\ref{gp_1000} and \ref{gp_1111}, the results from these two gauge choices are very similar.

\begin{figure}
\includegraphics[scale=.36, angle=-90]{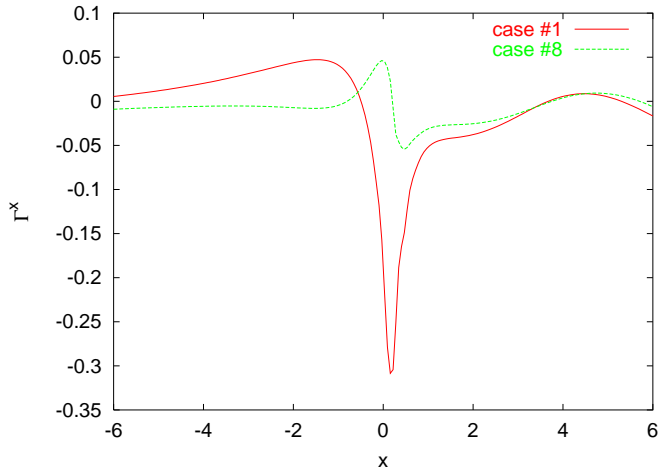}
\caption{${\tilde \Gamma}^x$ for the gauge ${\partial_t \alpha} = -2\alpha K + \beta^j\partial_j\alpha $,
$\partial_t \beta^i = \frac{3}{4}B^i$,
$\partial_t B^i = \partial_t{\tilde \Gamma}^i$ (Case \#1 with $\eta=0$) and for the gauge
$\partial_t \alpha = -2\alpha K + \beta^j\partial_j\alpha $,
$\partial_t \beta^i = \frac{3}{4}B^i + \beta^j\partial_j\beta^i$,
$\partial_t B^i = \partial_t{\tilde \Gamma}^i + \beta^j\partial_j(B^i-{\tilde \Gamma}^i)$
(Case \#8 with $\eta=0$)
at time $t=10M$. 
In the ``shifting-shift" case (Case \#8) ${\tilde \Gamma}^x$ is smaller
and smoother around the puncture than in the ``non-shifting-shift" case
(Case \#1).
}
\label{eta0}
\end{figure}

For these two gauge choices, we now consider varying the damping parameter, 
which has up to now been set to $\eta=2$.  By moving toward $\eta\to 0$
Eqs.~(\ref{eqn:shiftingshift_a}-\ref{eqn:shiftingshift_b}) appear a little closer to realizing a ``Gamma-freezing" 
condition.  Indeed, with $\eta=0$, we find that the the black holes 
come closer to realizing the physically expected velocity.
Both the ``shifting-shift" and ``non-shifting-shift" gauge options with $\eta=0$
have proven to allow stable evolutions of a single black hole.  
In Fig.~\ref{eta0} we find that in both cases setting $\eta=0$ results 
in significantly smaller values for 
${\tilde \Gamma}^i$, meaning a closer approximation to the Dirac gauge.
Further, we comment that we have observed numerically that for $\eta>0$, ${\tilde \Gamma}^i$ appears to show very slow linear growth in time,
whereas for $\eta=0$, ${\tilde \Gamma}^i$ appears to be bounded.
Comparing the ``shifting-shift" and ``non-shifting-shift" options with $\eta=0$ 
we now find more noticeable differences in ${\tilde \Gamma}^i$,
with our recommended ``shifting-shift" option giving a smoother result with ${\tilde \Gamma}^i$ holding closer to zero near the puncture.  

\begin{figure} 
\includegraphics[scale=.36, angle=-90]{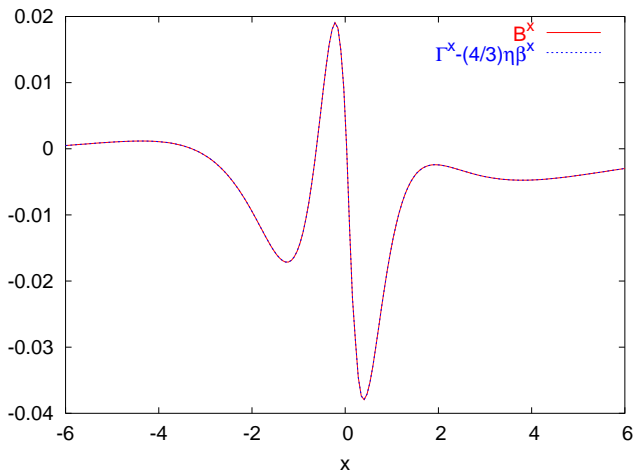} 
\caption{Results for $B^x$ and ${\tilde \Gamma}^x - \frac{4}{3}\eta\beta^x$ with the gauge ${\partial_t \alpha} = -2\alpha K + \beta^j\partial_j\alpha $,
${\partial_t \beta^i} = \frac{3}{4}B^i + \beta^j\partial_j\beta^i$,
${\partial_t B^i} = {\partial_t{\tilde \Gamma}^i} + \beta^j\partial_j(B^i-{\tilde \Gamma}^i) -\eta B^i $ at time $t=15M$.  The difference between the two
curves is within $10^{-11}$.
} 
\label{BGb} 
\end{figure}

\begin{figure}
\includegraphics[scale=.36, angle=-90]{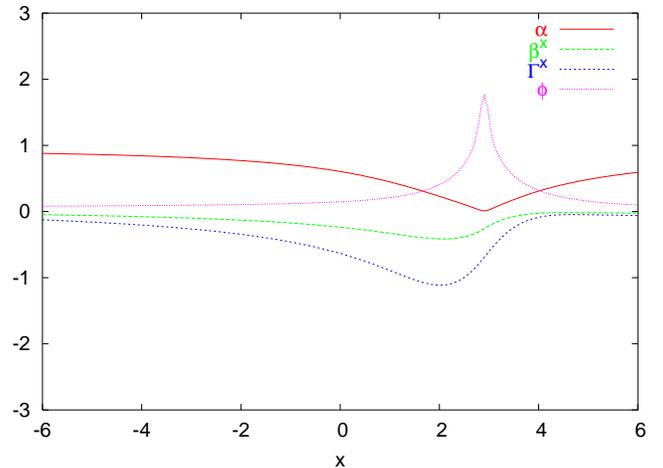}
\caption{Results for the gauge ${\partial_t \alpha} = -2\alpha K + \beta^j\partial_j\alpha $,
${\partial_t \beta^i} = \frac{3}{4}{\tilde \Gamma}^i  + \beta^j\partial_j\beta^i -\eta\beta^i$
at time $t=30M$.  The results are nearly identical to those depicted in
Fig.~\ref{gp_1111}
}
\label{gp_Bgone}
\end{figure}

These two gauges share an additional feature in common, which is that either Eqs.~(\ref{eqn:nonshiftingshift_a}-\ref{eqn:nonshiftingshift_b}) or Eqs.~(\ref{eqn:shiftingshift_a}-\ref{eqn:shiftingshift_b}) can be integrated to
give the relation
\begin{eqnarray} 
B^i = {\tilde \Gamma}^i - \frac{4}{3}\eta\beta^i 
\end{eqnarray}
since $B^i={\tilde \Gamma}^i=\beta^i=0$ initially.  This fact, also evident numerically (Fig.~\ref{BGb}), suggests substituting for 
$B^i$ in the evolution equation for $\beta^i$, to obtain (in the case of Eqs.~(\ref{eqn:shiftingshift_a}-\ref{eqn:shiftingshift_b}))
\begin{eqnarray} 
{\partial_t\beta^i} &=& \frac{3}{4}{\tilde \Gamma}^i  + \beta^j\partial_j\beta^i -\eta\beta^i
\label{eqn:Bgone}
\end{eqnarray}
Fig.~\ref{gp_Bgone} demonstrates that the resulting shift condition, Eq.~(\ref{eqn:Bgone}), can be evolved numerically to
yield the same stable, smooth simulation as with the analytically equivalent condition, Eqs.~(\ref{eqn:shiftingshift_a}-\ref{eqn:shiftingshift_b}), used previously.
Similar success should also be obtainable without the $\beta^j\partial_j\beta^i$ term.  Dispensing with the $B^i$ evolution equation results in a slightly more 
efficient numerical implementation, and also guarantees removal of the zero-speed mode associated with $B^i$.

Finally it may be noted that when $\eta=0$ and ${\tilde \Gamma}^i\to0$, Eq.~(\ref{eqn:Bgone}) (and Eqs.~(\ref{eqn:shiftingshift_a}-\ref{eqn:shiftingshift_b})) admits a ``shock" solution of the form 
$(x-x_0)/(t_0-t)$.  This represents the nonlinear tendency of this equation to
advect large magnitudes of the shift faster than smaller magnitudes, which for negative values and a positive
slope can lead to a vertical slope.  However, as $\partial_t{\tilde \Gamma}^i$ depends on derivatives of $\beta^i$,
${\tilde \Gamma}^i$ is unlikely to vanish when those derivatives become large.  Numerically $\beta^i$ has proven to be
very well behaved so ${\tilde \Gamma}^i$ evidently acts as an effective ``shock absorber", at the expense of not vanishing entirely.

\section{Conclusions}

We have investigated gauge conditions that are appropriate for the moving puncture approach to black hole simulations.
Potential hazards from zero-speed or slow-speed modes have been identified, a methodology for analytically exploring various gauge choices has
been presented, and a gauge free of slow-speed modes has been recommended.
Several gauges were also studied numerically and two gauges distinguished themselves as particularly well-adapted to smooth black hole motion.
In both cases we have suggested a simplification of the shift evolution equations, as well as explored the possibility of eliminating
a traditional damping term to better realize a ``Gamma-freezing" condition.  
We found that eliminating this damping term on the shift does indeed
minimize ${\tilde \Gamma}^i$ and yield smooth evolutions, more so with the
addition of our recommended ``shifting-shift" terms, and warrants further 
experimentation.  We intend to
investigate this $\eta\to0$ limit more thoroughly in future work.
\label{sec:conclusions}
\begin{acknowledgments}
This work was supported in part by NASA grants
ATP02-0043-0056 and O5-BEFS-05-0044.  The simulations were carried out using
Project Columbia at NASA Ames Research Center and at the NASA Center for
Computational Sciences at Goddard Space Flight Center. M.K and J.v.M.  were
supported by the Research Associateship Programs Office of the National
Research Council and the NASA Postdoctoral Program at the Oakridge
Associated Universities.  We are very grateful to Carsten Gundlach for
carefully reading and pointing out an error in an earlier draft.
\end{acknowledgments}
\bibliographystyle{../bibtex/apsrev}
\bibliography{../bibtex/references}
\end{document}